\documentclass{memsissa}
\usepackage[authoryear]{natbib}
\usepackage{txfonts}
\usepackage{balance}
\usepackage{graphicx}
\usepackage[a4paper]{hyperref}
\idline{75}{282}
\begin{document}
\def\teff{$T\rm_{eff }$}
\def\kms{$\mathrm {km s}^{-1}$}

\title{
Recent results on Pre-main sequence $\delta$ Scuti stars.
}

   \subtitle{}

\author{
V. \,Ripepi\inst{1} 
\and M. \, Marconi\inst{1}
\and F. \, Palla\inst{2}
\and S. \, Bernabei\inst{3,4}
\and A. \, Ruoppo\inst{1,5}
\and F. \, Cusano\inst{1,5}
\and J.M. \, Alcal\'a\inst{1}
}

  \offprints{V. Ripepi}

\institute{
Istituto Nazionale di Astrofisica --
Osservatorio Astronomico di Napoli, Via Moiariello 16,
I-80131 Napoli, Italy
\email{ripepi@na.astro.it,marconi@na.astro.it,jmea@na.astro.it}
\and
Istituto Nazionale di Astrofisica --
Osservatorio Astronomico di Arcetri, Largo E. Fermi 5,
I-50125 Firenze, Italy
\email{palla@arcetri.astro.it}
\and
Istituto Nazionale di Astrofisica --
Osservatorio Astronomico di Bologna, Via Ranzani 1, 
40127 Bologna, Italy 
\and
Departimento de Astrof\'{\i}sica, Universidad de La Laguna, Avda. 
Astrofisico F. S\'anchez sn, 30071 La Laguna, Spain
\email{stefano.bernabei@bo.astro.it}
\and 
Dipartimento di Scienze Fisiche, Universit\`a Federico II, Complesso 
Monte S. Angelo, 80126, Napoli, Italy
\email{ruoppo@na.astro.it,fcusano@na.astro.it} 
}

\authorrunning{Ripepi et al. }

\titlerunning{PMS $\delta$ Scuti stars}

\abstract{
Intermediate mass Pre-main sequence stars (1.5 $M_{\odot}< M < 5M_{\odot}$) 
cross the instability strip on their way to the main sequence. 
They are therefore
expected to be pulsating in a similar way as the $\delta$ Scuti stars. In this
contribution we present the status of the observational 
studies of pulsations in these stars with special emphasis on recent
results from our group.
The prospects for future investigations of these objects from the
ground and from space are discussed.
}
\maketitle{}

\section{Introduction}

Intermediate mass Pre-Main-Sequence (PMS) stars  
(M$>$1.5 M$_{\odot}$), often referred to as Herbig Ae/Be stars \citep{herbig},
are known to show photometric and spectroscopic variability on very different 
time-scales. Variable extinction due to circumstellar dust causes variations 
on week timescales, whereas clumped accretion or chromospheric activity is 
responsible for hours to days variability \citep[see e.g.][]{catala}. 
The fact that Herbig Ae/Be 
stars cross the pulsation instability strip of the more evolved 
$\delta$ Scuti stars during their contraction 
toward the Main Sequence, suggests that the variability 
on time scales of minutes to hours is due to pulsation 
\citep[see][]{baade,kurtz}.

Indeed the possible presence of pulsators among Herbig Ae/Be stars 
is particularly attractive since the precise observables which 
can be measured, i.e. the pulsation frequencies can, in principle, 
allow us to test evolutionary models by constraining the internal 
structure using asteroseismological techniques. 

The first detection of pulsation among Herbig Ae/Be stars date back to 
1972 with the discovery of  the two candidates V588 and V589 Mon in 
the young open cluster NGC 2264 \citep{breger1}. This initial finding was
confirmed by subsequent observations of  $\delta$ Scuti-like pulsations
in the Herbig Ae stars HR5999 \citep{kurtz} and HD104237 \citep{donati}.\par

These papers stimulated the first theoretical
investigation of the PMS instability strip based on non-linear
convective hydrodynamical models \citep{marconi} who calculated 
the topology of the PMS instability strip for the first three 
radial modes.
These authors also found that the interior structure of 
PMS stars crossing the instability strip is significantly different from 
that of more evolved Main Sequence stars (with the same mass and temperature), 
even though the envelopes structures are similar. 
This property was subsequently confirmed by Suran et al. (2001)
who made a comparative study of the seismology of a 1.8 $M_{\odot}$ 
PMS and post-MS star. Suran et al. (2001) pointed out that the 
unstable frequency range is approximately  the same for PMS and post-MS stars, 
but that some non-radial modes are very 
sensitive to the deep internal structure of the star. In particular, 
it is possible to discriminate between the PMS and post-MS stage, 
using differences 
in the oscillation frequency distribution in the low frequency 
range (i.e. $g$ modes).\\ 
%\citep[$g$ modes, see also][]{templeton}. \par
An other important aspect of PMS $\delta$ Scuti studies is represented by the 
possibility  to search for mode frequency 
changes due to evolution of the stellar inner structure 
\citep{pam}.
In fact, PMS evolutionary time scales are short enough to give 
relative variations of pulsation periods of the order of 
$\dot{P}/P$=10$^{-6}$, corresponding
to about 0.4 h in 10 years for the epoch of maxima in the 
case of the star HR 5999 \citep{catala}.

\section{Observations of PMS $\delta$ Scuti variables}

Since the seminal work by Marconi \& Palla (1998) our group started a 
systematic photometric monitoring  program of intermediate mass PMS 
stars with spectral 
types from A to F2-3 with the aims to: 1) identify the largest number of 
pulsating objects in order to observationally determine the boundaries of the 
instability strip for PMS $\delta$ Scuti pulsation; 2) study in detail 
 through multisite campaigns selected objects showing multiperiodicity 
\citep[see][]{marconi2001,h254,v346,v351,bernabei}.
The multiperiodic pulsators are potential candidates for future 
asteroseismological analysis.

Similar observational programs have been carried out
by various groups. As a results the current number of known or
suspected candidates amounts to about 34 stars 
\citep[see the updated list at http://ams.astro.univie.ac.at/pms\_corot.php, 
and the reviews by][]{zwintz,marconi2004,marconi2004a}. 
In particular over 34 candidates, 29 have been studied photometrically, but 
most of them have insufficient data due both to the short duration of 
the observations and/or to the poor duty cycle. Therefore most of the 
periodograms are affected by aliasing problems and are not useful for 
asteroseismology. \par

Only 5 stars have been observed by means of multisite campaigns: 
V588 and V589 Mon \citep[12 and 19 frequencies respectively, ][]{zwintz}, 
V351 Ori \citep[5 frequencies, ][]{v351}, IP Per \citep[9 frequencies, ][]
{ipper}, and HD 34282 \citep[9 frequencies, ][]{amado}.\par 
 
Concerning spectroscopic studies, radial velocities and line 
profile analysis (the latter being sensible to high degree modes, 
very useful for asteroseismology) have been carried out only for 
few stars: V351 Ori \citep[5 frequencies, ][]{balona}, $\beta$ Pic \citep[19 
frequencies, ][]{koen2}, HD 104237 \citep[5 frequencies, ][]{bohm}, 
and the binary star RS Cha \citep{alecian}. This last object is very 
interesting because it is an eclipsing double-lined spectroscopic 
binary. Preliminary results based on high resolution spectroscopy 
\citep{alecian} seem to show that both components are pulsating. 
This star therefore will offer the unique opportunity to obtain 
stringent constraints on pulsating models. \par

\section{Theoretical interpretation}

\begin{figure}[t!]
%\centering
\resizebox{6.5cm}{!}{\includegraphics[clip=true]{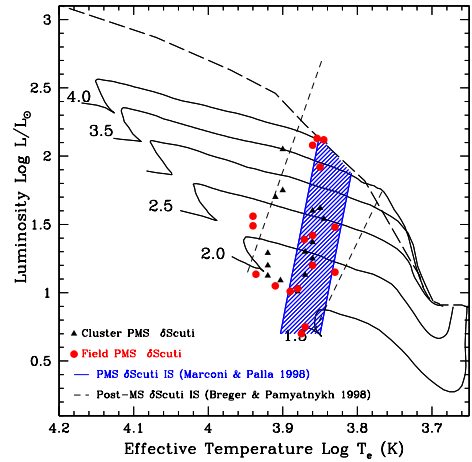}}
\caption{\footnotesize
The position of PMS $\delta$ Scuti stars in the HR diagram as predicted on the 
basis of the comparison between the observed periodicities and linear 
nonadiabatic radial pulsation models. The shaded region is the 
theoretical instability strip for the first three radial modes \citep{marconi}, 
that is the region between the second overtone blue edge and the 
fundamental red edge. The dashed lines represent the instability strip 
of more evolved $\delta$ Scuti stars \citep{pam}. 
}
\label{fig1}
\end{figure}

The comparison between observed frequencies and those predicted 
by linear non-adiabatic pulsation analysis allows us to evaluate 
the position in the HR diagram and the mass for both field and 
cluster pulsator (see Fig.~\ref{fig1}).
No object is predicted to be located to the right of the red 
boundary of the theoretical instability strip by Marconi \& Palla (1998).
Some objects are predicted to pulsate in higher overtones
than the second one and then to be located to the left of 
the 2$^{nd}$ overtone blue edge. However it can be seen in Fig.~\ref{fig1}) 
that the majority of the pulsators are located within the observed 
instability strip for evolved $\delta$ Scuti stars \citep{pam}. 

Most of the well observed candidates PMS $\delta$ Scuti stars show 
frequencies which cannot be reproduced by radial analysis only. Clearly 
non-radial modes are present in this class of stars. A thorough interpretation 
of observed frequencies at the light of non-radial pulsation theory is still 
lacking for  PMS $\delta$ Scuti stars. 

\section{The case of IP Per}

\begin{figure}[t!]
\centering
\resizebox{6.5cm}{!}{\includegraphics[clip=true]{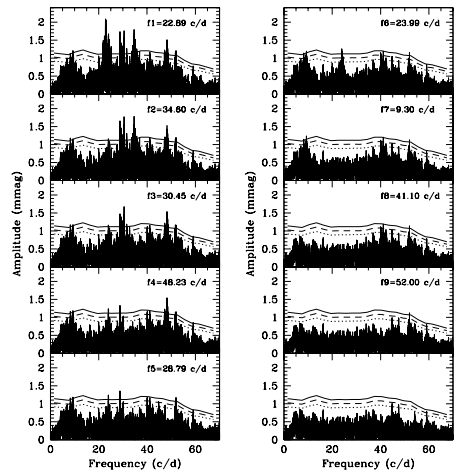}}
\caption{\footnotesize Frequency analysis for the V dataset. The solid, 
dashed and dotted lines show respectively the 99.9\%, 99\% and 90\% 
significant levels determined in a standard way \citep{kus97}.}
\label{per}
\end{figure}

IP Per is a Herbig Ae star with: V=10.34 mag, spectral type A7V, 
$\log L/L_{\odot}\sim1.0\pm0.05$ dex,
\teff$\sim$8000$\pm$200 K \citep{miro}. By using this physical parameters
in the HR diagram, IP Per falls in the instability strip for $\delta$ Scuti pulsation by
Marconi \& Palla (1998). In order to study in detail this star, we carried out 
a multisite campaign involving nine telescopes around the world 
\citep[for details see][]{ripepiiau,ipper}. 
As a result we were able to detect nine frequencies of pulsation as 
shown in Fig.~\ref{per} where we report the Fourier Frequency analysis 
for the visual datatset.

\begin{figure}[t!]
\centering
\resizebox{6.5cm}{!}{\includegraphics[clip=true]{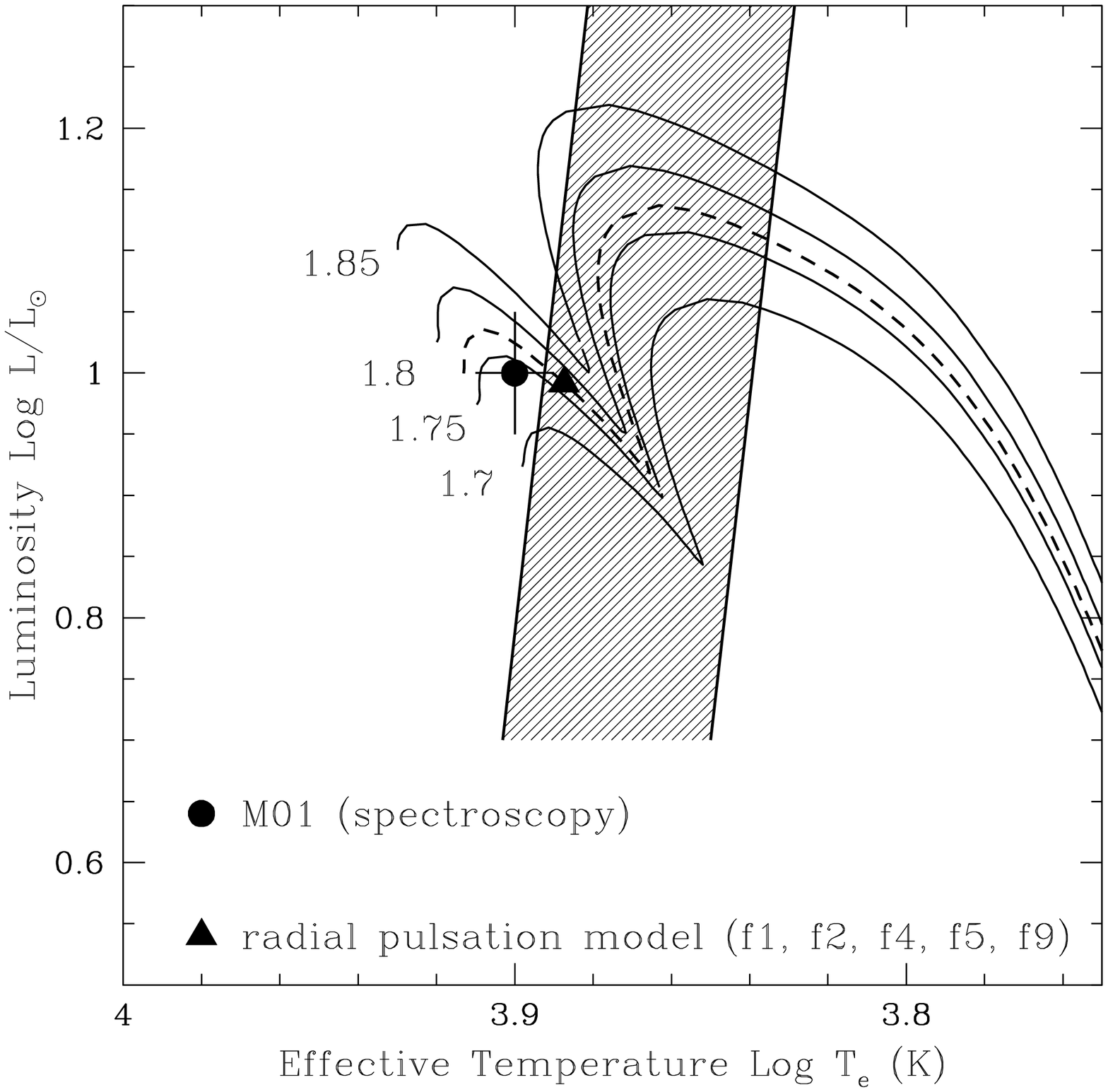}}
\caption{\footnotesize 
Position in the HR diagram of the best-fit radial pulsation model for IP Per. 
The shaded region is the predicted instability strip by 
Marconi \& Palla (1998). 
Solid lines are the PMS evolutionary tracks computed for the 
labelled stellar masses with the FRANEC stellar evolution code 
\citep{chieffi,castellani}. The dashed lines 
represent the 1.77$M_{\odot}$ PMS track with solar composition. 
The empirical determination based on the spectroscopic measurements \citep{miro} 
is shown with a filled circle.}
\label{hr}
\end{figure}

The observed periodicities can be used to constrain the 
intrinsic stellar properties of IP Per and in particular its mass and position 
in the HR diagram, through comparison with stellar pulsation models.
Using a linear non-adiabatic pulsation code \citep[see][]{marconi,marconi2004a} 
we could not reproduce all the observed
frequencies. In fact, we can recover at most 5  of the 9 
observed frequencies for $M=1.77\pm0.01 M_{\odot}$, 
$\log{L/L_{\odot}}=0.992\pm0.003$,
$\log T_{\rm eff}=3.887\pm0.002$. This solution corresponds to a radial 
pulsation model which simultaneously oscillates in the first ($f1$), 
second ($f5$), third ($f2$), fifth ($f4$) and sixth ($f9$)
overtones. Its position  in the HR diagram is shown in Fig.~\ref{hr} together 
with the predicted instability strip by Marconi \& Palla (1998) and the PMS 
evolutionary tracks computed for the labelled stellar masses with 
the FRANEC stellar evolution code \citep{chieffi,castellani}.  
The $1.77 M_{\odot}$ PMS track is represented by the dotted line.
Note that the predicted position in the HR diagram is consistent with the 
empirical determination based on the spectroscopic measurements  
\citep{miro} represented by the filled circle in the figure.

Non-radial pulsation is clearly also present in this star. A preliminary 
interpretation of the observed frequencies through the Aarhus 
adiabatic non-radial pulsation code (http://astro.phys.au.dk/$\sim$jcd/adipack.n/), 
applied to the evolutionary structure of the  1.77 $M_{\odot}$ 
model reproducing $f1$, $f2$, $f4$, $f5$ and $f9$ with radial modes, 
seems to indicate that $f3$, $f6$ and $f8$  are associated with non-radial
modes with $l=2$.

\section{Future prospects}

The future of PMS $\delta$ Scuti studies relies mainly    
on space missions. Indeed, two asteroseismological satellites have 
the possibility to observe PMS stars: MOST and COROT. 
MOST (Microvariability and Oscillations of STars, http://www.astro.ubc.ca/MOST/) 
is equipped with a 15 cm telescope and a CCD camera. MOST already observed 
the two prototypes of the class: V588 and V589 Mon. The results will 
be available in the near future. \par
\begin{figure}[t!]
\centering
\resizebox{7cm}{!}{\includegraphics[clip=true]{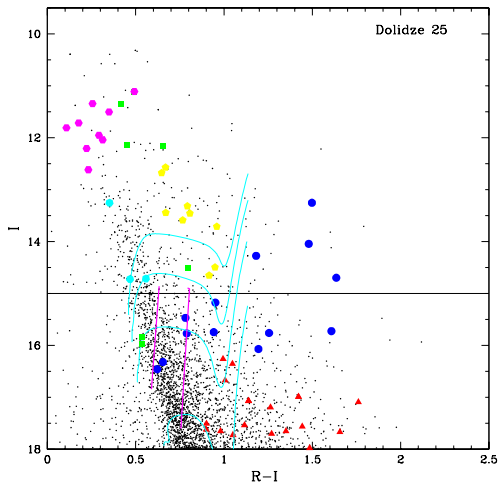}}
\caption{\footnotesize CM-diagram in the I vs (R-I) plane for an area 
of about 25$^{'}$x25$^{'}$ centered on Dolidze 25 as obtained from 
VIMOS@VLT pre-imaging data (black dots). 
The various symbols (except the red triangles) show the young stars proposed 
to be observed with COROT. Note that at least four object fall within the 
observed instability strip for PMS $\delta$ Scuti pulsation 
(magenta solid line).
Cyan lines show PMS tracks for Z=0.008 and masses of 1.5,2.5,3.5 and 4.5 
M$_{\odot}$ (Degl'Innocenti \& Marconi, private communication). 
As reference, the black solid line show roughly the level of V=16 mag}
\label{do25}
\end{figure}

   
COROT (COnvection and ROTation of stars; http://corot.oamp.fr) will be 
launched in June 2006 and will observe a few fields for 5 months 
continuously and a few others for 10-20 days. In the context of 
Additional Programs there are a few proposals  
having also PMS $\delta$ Scuti stars as targets. Among these projects, we 
would describe more in detail the one (already approved by SC, P.I. V. Ripepi) 
concerning the observation (in the ESO-Planet field) of  few selected 
stars in Dolidze 25, a distant, metallicity-deficient young open 
cluster which fall in the continuous viewing zone of COROT. 
In order to identify PMS objects in this cluster, we 
carried out a photometric (RI filters) and spectroscopic investigation 
of Dolidze 25 by using VIRMOS@VLT instrument. Photometry has been used 
to select targets for spectroscopy. In total we obtained $\sim$900 spectra with 
medium resolution (2.5\AA/pixel) and $\sim$600 with 
higher resolution (0.6\AA/pixel).
The result of this investigation is reported in Fig.~\ref{do25}, where we 
present the CMD in the I vs (R-I) plane.
The blue filled circles in the figure identify all the objects 
with R$\leq$17.2 for which we have found H$\alpha$ emission. 
As shown in the figure at least four of these objects fall within 
or near the observed instability strip 
for PMS pulsation.

\section{Conclusions}

Asteroseismology applied to PMS $\delta$ Scuti stars would allow us 
to test the evolutionary status and the internal structure of these 
objects. However more theoretical work is needed in order to interpret 
present observations. \par
Observationally we are still in an early phase: the empirical 
instability strip is not well known and only for few stars the derived  
frequency spectrum is accurate enough to use  asteroseismological 
techniques.\par
We expect great improvements in the study of PMS $\delta$ Scuti stars 
from space observations with the satellites MOST and COROT.

%\begin{acknowledgements}
%\end{acknowledgements}

\bibliographystyle{aa}

\end{document}